\documentclass[aps,twocolumn,pra,superscriptaddress,showpacs,tightenlines]{revtex4}
\usepackage{amssymb}
\usepackage{amsmath}
\usepackage{graphicx}
\usepackage{epsfig}

\begin{document}

\title{Spin entanglement induced by spin-orbit interactions in coupled
quantum dots}
\author{Nan Zhao}
\affiliation{Department of Physics, Tsinghua University, Beijing 100084, China}
\author{L. Zhong}
\affiliation{Institute of Theoretical Physics, Chinese Academy of Sciences, Beijing,
100080, China}
\author{Jia-Lin Zhu}
\affiliation{Department of Physics, Tsinghua University, Beijing 100084, China}
\author{C. P. Sun}
\affiliation{Institute of Theoretical Physics, Chinese Academy of Sciences, Beijing,
100080, China}

\begin{abstract}
We theoretically explore the possibility of creating spin quantum
entanglement in a system of two electrons confined respectively in two
vertically coupled quantum dots in the presence of Rashba type spin-orbit
coupling. We find that the system can be described by a generalized Jaynes -
Cummings model of two modes bosons interacting with two spins. The lower
excitation states of this model are calculated to reveal the underlying
physics of the far infrared absorption spectra. The analytic perturbation
approach shows that an effective transverse coupling of spins can be
obtained by eliminating the orbital degrees of freedom in the large detuning
limit. Here, the orbital degrees of freedom of the two electrons, which are
described by two modes of bosons, serve as a quantized data bus to exchange
the quantum information between two electrons. Then a nontrivial two-qubit
logic gate is realized and spin entanglement between the two electrons is
created by virtue of spin-orbit coupling.
\end{abstract}

\maketitle

\section{Introduction}

Control and manipulation of the spin degree of freedom become one of the
most important topics both in spintronics\cite{SpintroincsReview} and in
quantum information processing \cite{qunt-infor Book}. The spin-orbit
interactions (SOI) in semiconductor heterostructures provides the ways to
couple spin with spatial degree of freedom, and consequently has attracted
more and more attention in recent years. The spin properties of a few
electrons confined in semiconductor nanostructures, such as quantum dots\cite%
{Debald2004, Governale2002, Lucignano2005, Chakraborty2005,
Chakraborty2005Arxiv,ChakrabortyPara}, coupled quantum dots\cite{Stano},
quantum rings\cite{Splettstoesser} and quantum wires\cite{Debald2005}, have
been studied. The results show that the carrier spin properties are strongly
affected in the presence of the SOI, and novel features emerge in these
nonostructures compared with the traditional ones without SOI.

On the other hand, spin confined in quantum dot is a natural choice for the
physical realization of qubit. This kind of system is considered as an
important candidate for solid state based quantum computing. Among various
approaches to implement quantum information processing using quantum dot
systems, the optical method, including the classical laser field\cite{ZhuJL}
and the quantized cavity modes \cite{sham}, has been proposed to create
entanglement and to realize the single and double qubit logic gate. In Ref.
\cite{Imamoglu}, a scheme with quantum dots embedded in an optical cavity
was designed, so that the cavity mode can serve as a data bus and induce a
spin-spin interaction. This kind of cavity-mediated two-qubit gate is
studied in details for several other solid state systems very recently\cite%
{DLoss}.

Now we notice that the SOI phenomena in nanostructures were investigated
with the help of the quantum optics method. Taking advantage of the
tunability of the SOI strength, an experiment to observe coherent
oscillations in a single quantum dot was proposed in ref. \cite{Debald2004}.
In this proposal, the orbital degrees of freedom are modeled by two boson
modes. Under the rotating wave approximation (RWA), the SOI of the electron
confined in the single quantum dot, is reduced to a Jaynes- Cummings (JC)
model, which is very typical model in quantum optics. This analogy between
the SOI in a quantum dot and the JC model in cavity QED suggests us that it
is possible to make use of the orbital degree of freedom, instead of the
real optical cavity modes, to serve as a quantized data bus and then to
induce a spin entanglement\cite{suncp}.

In this paper, we propose and study a model of two electrons confined in two
quantum dots respectively with Rashba type SOI, and explore the possibility
of realizing a two-qubit logic gate or creating spin entanglement with this
system. For simplicity, we consider two vertically coupled quantum dots
(CQDs) with two dimensional parabolic confinements. In the case with strong
confinement and large interdot separation, the Coulomb interaction between
the two electrons is approximately expanded in a quadratic form, and then
the orbital part of the two electron system can be reduced to four-mode
bosons. Under the RWA, only two of the four modes are coupled with the spin
degrees of freedom. Then it is proved that the total system with SOI can be
effectively described by a generalized JC model with the coupling between
two bosons and two spins. By diagonalizing the Hamiltonian directly in the
lower excitation subspace, the eigenvalues and the corresponding
eigenfunction are obtained exactly. The far infrared (FIR) absorption
spectra are calculated according to these analytical solutions, which help
us to understand the underlying physics of the spectra. To get an effective
Hamiltonian of spin-spin interaction, we perform the Fr{\"{o}}hlich
transformation in the large detuning limit. This effective Hamiltonian can
dynamically drive a two-qubit logic gate operation. By using the
conventional material parameters, our numerical estimation shows that the
effective spin interaction induced by SOI is strong enough, compared to the
spin decoherence in low dimensional semiconductor structures. It is feasible
experimentally to implement a two-qubit logic gate and thus produce quantum
entanglement.

The paper is organized as follows. In Section II, the two modes JC model is
derived from the conventional Hamiltonian of two electrons confined in two
quantum dots. Analytical solution of lower excitation states and FIR spectra
are presented in Section III. In Section IV, we show that the perturbation
treatment gives the effective transverse spin-spin interaction in the large
detuning limit, and demonstrate that a two-qubit logic gate and quantum
entanglement can be achieved in this kind system with SOI.
\begin{figure}[tbp]
\includegraphics[bb=55 130 540 700, width=7 cm, clip]{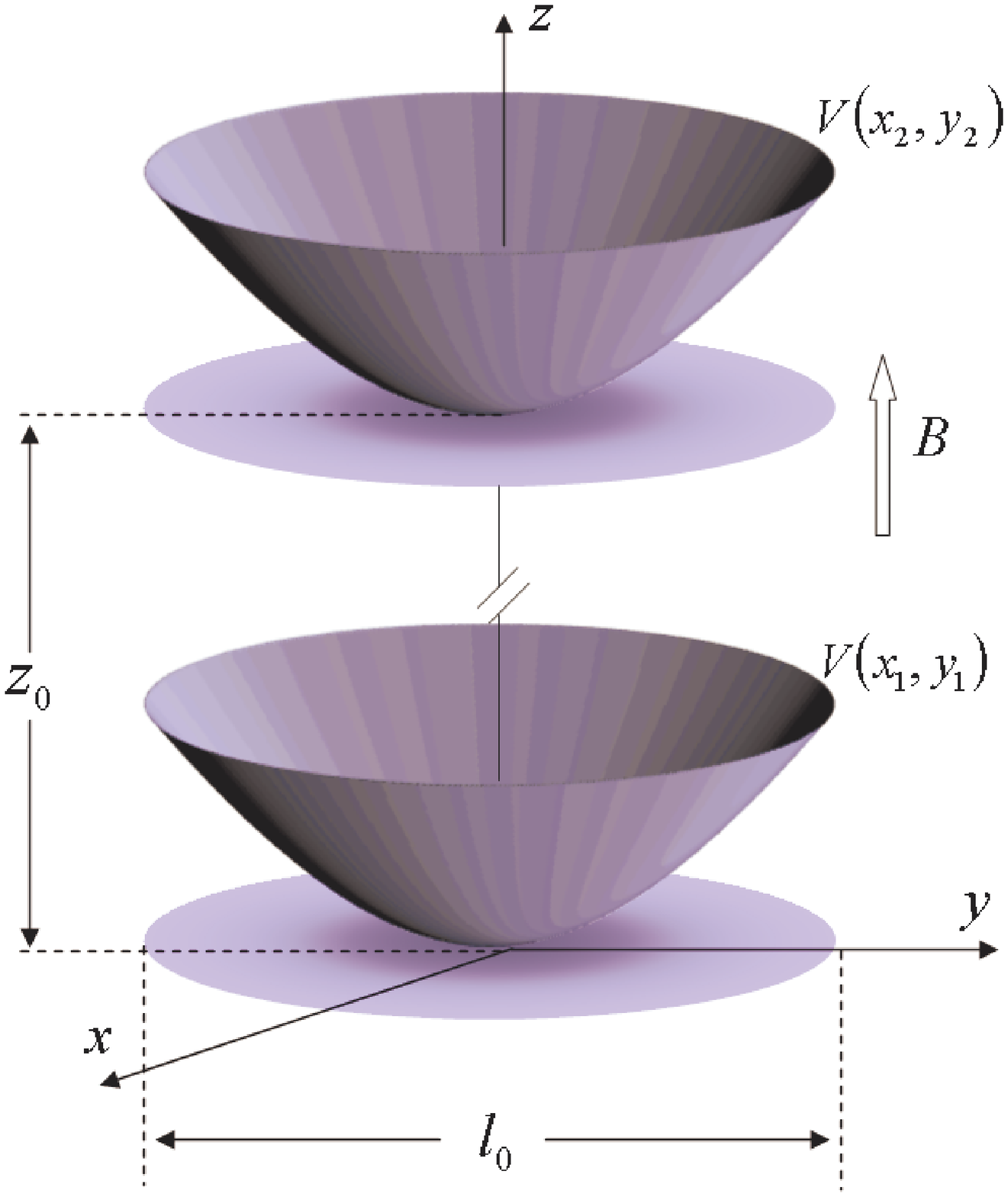}
\caption{{}Schematic illustration of the vertically stacked quantum dots.
Each quantum dot confined an electron. We assume that the interdot
separation $z_{0}$ is larger than the characteristic length of quantum dot
confinement $l_{0}$. }
\label{Scheme}
\end{figure}

\section{Two modes JC model for vertically CQDs with SOI}

We consider two vertically CQDs with an interdot separation $z_{0}$, see Fig.%
\ref{Scheme}. Each quantum dot is described by a two dimensional parabolic
confinement potential $V\left( r_{i}\right) =m_{0}\omega _{0}^{2}r_{i}^{2}/2$
with the basic frequency $\omega _{0}.$ Including the Rashba type SOI
\begin{equation}
H_{SO}^{\left( i\right) }=\frac{\alpha }{\hbar }\left[ \mathbf{\sigma }%
_{i}\times \mathbf{\Pi }_{i}\right] _{z},
\end{equation}%
the Zeeman term
\begin{equation}
H_{Z}^{\left( i\right) }=\frac{1}{2}g\mu _{B}\mathbf{B\cdot \sigma }_{i},
\end{equation}%
and the Coulomb interaction
\begin{equation}
V_{coul}(\left\vert \mathbf{r}_{1}-\mathbf{r}_{2}\right\vert )=\frac{e^{2}}{%
4\pi \varepsilon \varepsilon _{0}\left\vert \mathbf{r}_{1}-\mathbf{r}%
_{2}\right\vert },
\end{equation}
the total Hamiltonian reads:%
\begin{equation}
H=\sum_{i=1,2}\left( \frac{\mathbf{\Pi }^{2}}{2m_{0}}+V\left( r_{i}\right)
+H_{SO}^{\left( i\right) }+H_{Z}^{\left( i\right) }\right) +V_{coul}.
\end{equation}%
Here, $-e$ , $\mu _{B}$ and $\varepsilon _{0}$ are the electron charge, Bohr
magneton and dielectric constant in vacuum, $m_{0}$, $g$ and $\varepsilon $
are the material related parameters of effective mass, Landre g-factor and
the relative dielectric constant, respectively. $\mathbf{\Pi }_{i}=\mathbf{p}%
_{i}+e\mathbf{A}\left( \mathbf{r}_{i}\right) $ is the canonical momentum and
$\mathbf{A}(\mathbf{r}_{i})=B\left( -y_{i}/2,x_{i}/2,0\right) $ is the
vector potential for magnetic field $\mathbf{B}=B\mathbf{\hat{z}}$.

In order to simplify the Coulomb interaction, we consider a special case in
which the interdot separation $z_{0}$ is much larger than the lateral
confinement characteristic length $l_{0}=\sqrt{\hbar /m_{0}\omega _{0}}$,
i.e. $\left( l_{0}/z_{0}\right) ^{2}<<1$. Then we expand the Coulomb
interaction as a power series of the relative coordinate $r=\left\vert
\mathbf{r}_{1}-\mathbf{r}_{2}\right\vert $ up to the second order\cite%
{CoulombExpandPRL}\cite{CoulombExpandPRB}:%
\begin{equation}
V_{coul}\left( r\right) \simeq V_{0}-\frac{1}{2}m_{0}\omega _{1}^{2}r^{2}.
\end{equation}%
Here, we have defined $V_{0}=e^{2}/\left( 4\pi \varepsilon \varepsilon
_{0}z_{0}\right) $, and $\hbar \omega _{1}=\sqrt{\hbar ^{2}V_{0}/mz_{0}^{2}}$%
. We also assume that electrons are strictly confined in each quantum dots
planes, and then neglect the overlap of their wavefunctions.

In the center of mass (CM) reference of frame defined by $\mathbf{R}=\left(
\mathbf{r}_{1}+\mathbf{r}_{2}\right) /2$ and $\mathbf{r}=\mathbf{r}_{1}-%
\mathbf{r}_{2},$we have the C.M \ and relative momentums, and the
corresponding angular momentums $\mathbf{P}=\mathbf{p}_{1}+\mathbf{p}_{2},%
\mathbf{p}=\frac{1}{2}\left( \mathbf{p}_{1}-\mathbf{p}_{2}\right) ,\mathbf{L}%
=\mathbf{R}\times \mathbf{P},$ and $\mathbf{l}=\mathbf{r}\times \mathbf{p}$,
where $M=2m_{0},m=m_{0}/2,\mathbf{r}$ is the relative coordinate and $%
\mathbf{R}$ the CM coordinate. The orbital part of the Hamiltonian is
expressed in a quadrature form of these coordinates:
\begin{eqnarray}
H_{orbit} &=&\sum_{i=1,2}\left( \frac{\mathbf{\Pi }_{i}^{2}}{2m_{0}}+V\left(
r_{i}\right) \right) +V_{coul}  \notag  \label{HamiOrbit} \\
&=&\frac{P^{2}}{2M}+\frac{1}{2}M\Omega ^{2}R^{2}+\frac{1}{2}\omega _{c}L_{z}
\notag \\
&+&\frac{p^{2}}{2m}+\frac{1}{2}m\omega ^{2}r^{2}+\frac{1}{2}\omega _{c}l_{z}
\end{eqnarray}%
Here, the cyclone frequency is $\omega _{c}=eB/m_{0}$, and frequency of CM
and relative motion $\Omega =\sqrt{\omega _{0}^{2}+\omega _{c}^{2}/4}$ and $%
\omega =\sqrt{\Omega ^{2}-2\omega _{1}^{2}}$, respectively. Note that the
effect of the Coulomb repulsion is reducing the relative motion to a lower
frequency compared to the CM motion. In our model the requirement $\left(
l_{0}/z_{0}\right) ^{2}<<1$ ensures that the $\omega >0$ is satisfied even
when $B=0$.

\begin{figure}[tbp]
\includegraphics[bb=20 80 550 775, width=7 cm, clip]{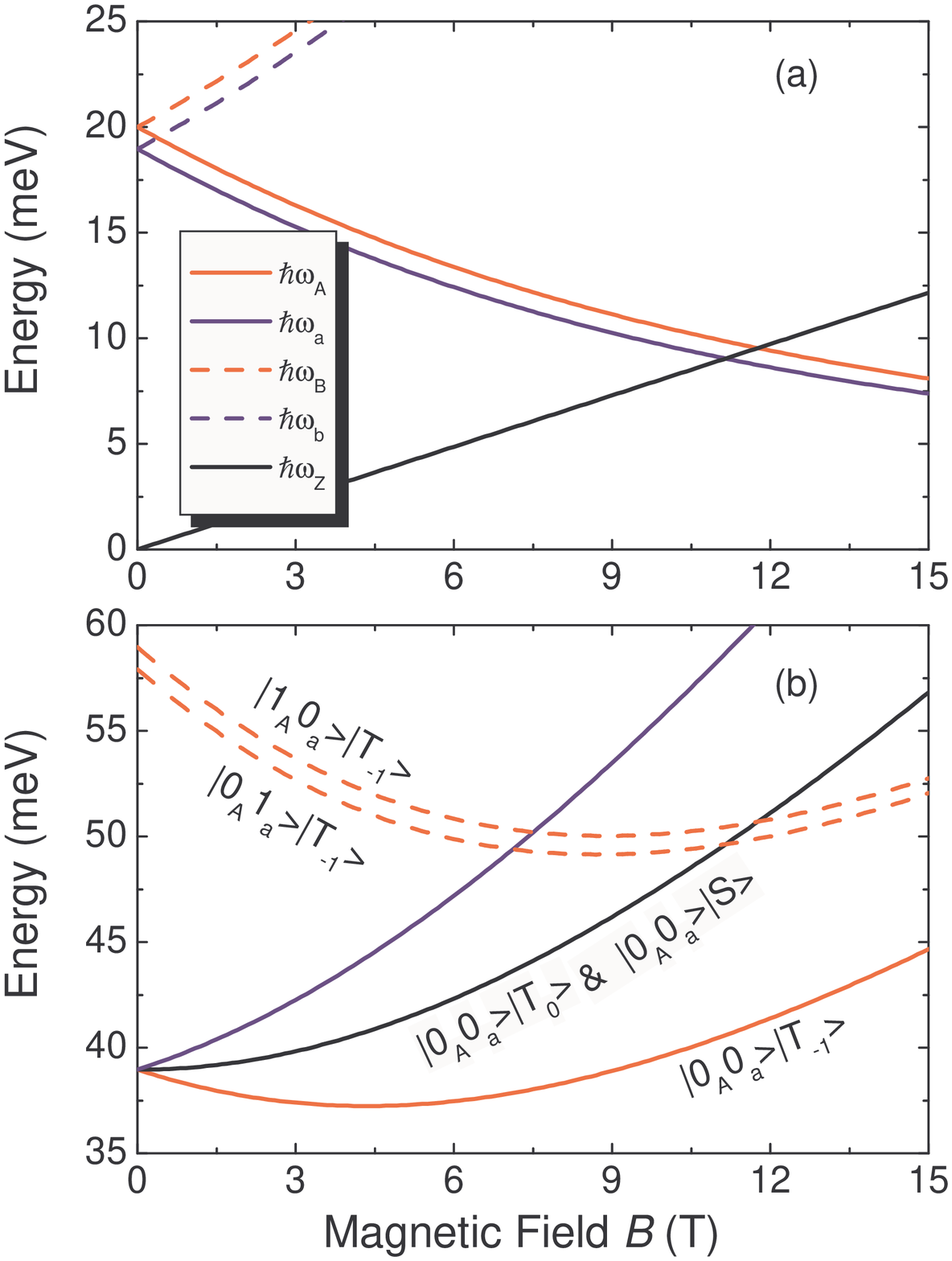}
\caption{(a){} Energy dispersions with respect to the magnetic field $B$ of
the four modes of the boson frequencies $\protect\omega _{A}$ (red solid
line), $\protect\omega _{a}$ (blue solid line), $\protect\omega _{B}$ (red
dashed line), and $\protect\omega _{b}$ (blue dashed line), and the Zeeman
energy $\protect\omega _{z}$ (black line). (b) Energy spectrum of $H_{0}$.
Different lines correspond to different spin and orbit states, which are
indicated explicitly in the figure. The parameters used in calculation are
listed in Table \protect\ref{ParaValueTable}}
\label{nonRashba}
\end{figure}

We define the boson operators $a_{x},a_{x}^{+},a_{X},a_{X}^{+}$ of \ $x$
components by
\begin{align}
X& =\sqrt{\frac{\hbar }{2M\Omega }}\left( a_{X}^{+}+a_{X}\right) ,  \notag \\
x& =\sqrt{\frac{\hbar }{2m\omega }}\left( a_{x}^{+}+a_{x}\right) ,  \notag \\
P_{X}& =i\sqrt{\frac{\hbar M\Omega }{2}}\left( a_{X}^{+}-a_{X}\right) ,
\label{bosonOperators} \\
p_{x}& =i\sqrt{\frac{\hbar m\omega }{2}}\left( a_{x}^{+}-a_{x}\right)  \notag
\end{align}%
and the boson operators $a_{y},a_{y}^{+},a_{Y},a_{Y}^{+}$ of $y$ components
\ are defined in the same way. Let%
\begin{align}
A& =\left( a_{X}+ia_{Y}\right) /\sqrt{2},  \notag \\
a& =\left( a_{x}+ia_{y}\right) /\sqrt{2},  \notag \\
B& =\left( a_{X}-ia_{Y}\right) /\sqrt{2},  \label{bosonOperators2} \\
b& =\left( a_{x}-ia_{y}\right) /\sqrt{2}.  \notag
\end{align}%
\begin{table}[tbp]
\begin{center}
\begin{tabular}{ccc|ccc}
\hline\hline
Quantity & Value & Unit & Quantity & Value & Unit \\ \hline
$m_{0}$ & 0.042 & - & $\omega _{0}$ & 20 & meV \\
$\varepsilon $ & 14.6 & - & $l_{0}$ & 9.5 & nm \\
$g$ & -14 & - & $z_{0}$ & 20 & nm \\
$\alpha $ & 10 & meV$\cdot $nm &  &  &  \\ \hline\hline
\end{tabular}%
\end{center}
\caption{Parameters used in the calculations}
\label{ParaValueTable}
\end{table}
Then we rewrite orbit Hamiltonian (\ref{HamiOrbit})
\begin{align}
H_{\text{orbit}}& =\hbar \omega _{A}\left( A^{+}A+1/2\right) +\hbar \omega
_{B}\left( B^{+}B+1/2\right)  \notag \\
& +\hbar \omega _{a}\left( a^{+}a+1/2\right) +\hbar \omega _{b}\left(
b^{+}b+1/2\right)  \label{HamiFourmodes}
\end{align}%
in terms of four mode bosons $A,a,B,$and $b$, with their frequencies
respectively%
\begin{align}
\omega _{A}& =\Omega -\frac{1}{2}\omega _{c},\omega _{B}=\Omega +\frac{1}{2}%
\omega _{c},  \label{BosonFreq1} \\
\omega _{a}& =\omega -\frac{1}{2}\omega _{c},\omega _{b}=\omega +\frac{1}{2}%
\omega _{c}  \label{BosonFreq2}
\end{align}%
These four frequencies, together with the Zeeman energy $\hbar \omega
_{z}=\left\vert g\right\vert \mu _{B}B$ are drawn in Fig.\ref{nonRashba}
with respect to the magnetic field $B$. Note that the Zeeman energy $\hbar
\omega _{z}$ and the boson frequencies $\omega _{A}$ and $\omega _{a}$ reach
the resonant regime at $B\simeq 11T$ with the parameters listed in Table \ref%
{ParaValueTable}. We also draw the lower energy spectra of the orbit
Hamiltonian (\ref{HamiOrbit}) in Fig.\ref{nonRashba}(b). We will focus on
how these states are affected in the presence of SOI in the following
sections.

In terms of the four boson operators defined by Eqs. (\ref{bosonOperators})
and (\ref{bosonOperators2}), the SOI Hamiltonian $H_{SO}$, after some
straightforward algebra, can be rewritten as:%
\begin{eqnarray}
H_{SO}^{RWA} &=&g_{A}\cdot A\left( \sigma _{1+}+\sigma _{2+}\right)  \notag
\\
&+&g_{a}\cdot a\left( \sigma _{1+}-\sigma _{2+}\right) +h.c.
\end{eqnarray}%
Note that, due to the negative value of the Landre g-factor, a unitary
rotation $\sigma _{z}\mapsto -\sigma _{z}$ and $\sigma _{\pm }\mapsto
-\sigma _{\mp }$ has been performed. To obtain the interaction Hamiltonian
above, we have used the RWA to neglect the counter-rotating terms like $%
\sigma _{+}B^{+},\sigma _{+}b^{+},\sigma _{-}B,$and $\sigma _{-}b$. This
approximation has been verified numerically for single electron case in Ref.
\cite{Debald2004}. The coupling strengths $g_{A}$ and $g_{a}$ are defined as
follows:%
\begin{align}
g_{A}& =\alpha \sqrt{\frac{m_{0}\Omega }{2\hbar }}\left( 1-\frac{\omega _{c}%
}{2\Omega }\right)  \label{CouplingStrengthA} \\
g_{a}& =\alpha \sqrt{\frac{m_{0}\omega }{2\hbar }}\left( 1-\frac{\omega _{c}%
}{2\omega }\right)  \label{CouplingStrengtha}
\end{align}

So far, we have obtained a generalized JC model where two-mode bosons
interact with two spins. In the following sections we will further
demonstrate how the orbit motion induce a spin-spin entanglement in the
presence of SOI.

\section{Lower excitation states and FIR spectra}

\begin{figure}[tbph]
\includegraphics[bb=15 60 555 755, width=7 cm, clip]{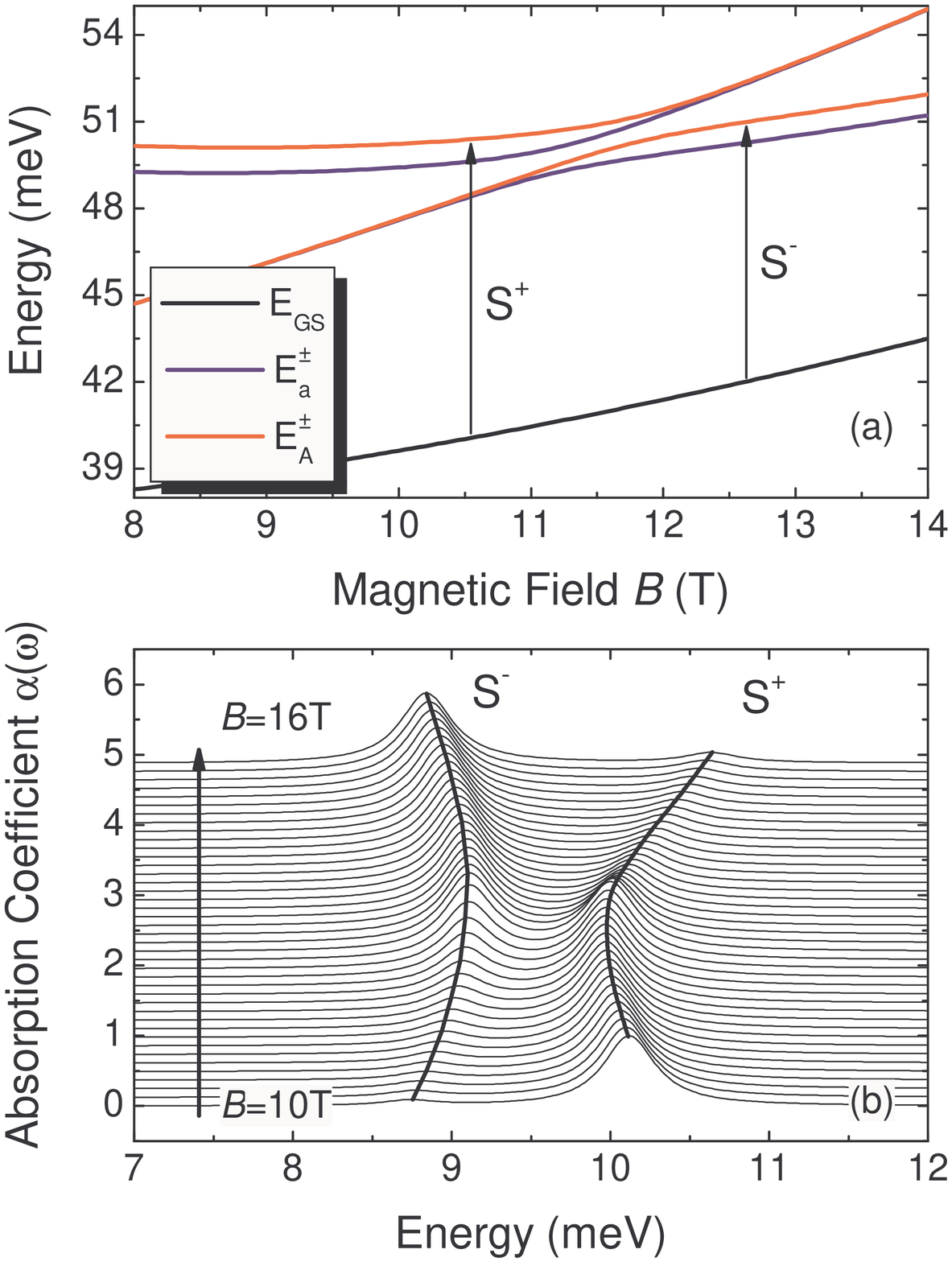}
\caption{{}(a) Exact energy spectra given by Eqs.(\protect\ref{ExactEnergy1}%
) and (\protect\ref{ExactEnergy}). The arrows indicate the dipole allowed
transition from $\left\vert GS\right\rangle $ (black line) to $\vert \Phi
_{A}^{\pm }\rangle $ (red lines). $\vert \Phi _{a}^{\pm }\rangle $ (blue
lines) is dipole inactive. (b) FIR spectra for increasing magnetic field $B$%
. The anti-crossing is clearly shown. A Lorentz profile function $\Gamma /%
\protect\pi [ ( \protect\omega -\protect\omega _{fi}) -\Gamma ^{2}] $ is
used to replace the $\protect\delta $ function. The phenomenological broaden
factor $\Gamma =0.2meV.$ Spectra are normalized to their maxima and offset
for clarity.}
\label{EnergySpctra}
\end{figure}

In this section, we calculate the eigenenergies and eigenstates of the low
excitation states. Notice that the total excitation number operator
\begin{equation}
\hat{N}=a^{+}a+A^{+}A+\frac{1}{2}\left( \sigma _{1z}+\sigma _{2z}\right)
\end{equation}%
commutes with $H=H_{\text{orbit}}+H_{Z}^{\left( 1\right) }+H_{Z}^{\left(
2\right) }+H_{SO}^{RWA}$. For a given integer $N$, which is the eigenvalue
of $\hat{N}$, the dimension of the invariant subspace $V^{\left( N\right) }$
is $4N+4$ for $N\geq 0$.

The lowest subspace $V^{\left( -1\right) }$, corresponding to $N=-1$, is of
one dimension. The the ground state can be directly written as $\left\vert
GS\right\rangle =\left\vert 0_{A},0_{a},0_{B},0_{b},\downarrow ,\downarrow
\right\rangle $, which means the excitation numbers of boson modes
\begin{equation}
A^{+}A=a^{+}a=B^{+}B=b^{+}b=0
\end{equation}
and both spins are in the down state. The corresponding eigenenergy is
\begin{equation}
E_{GS}=\frac{\hbar }{2}\left( \omega _{A}+\omega _{B}+\omega _{a}+\omega
_{b}-2\omega _{z}\right) .
\end{equation}

The second lowest subspace $V^{\left( 0\right) }$, which is of four
dimension. The Hamiltonian can be exactly diagnolized in this subspace. We
define
\begin{eqnarray}
\left\vert S\right\rangle &=&\frac{1}{\sqrt{2}}\left( \left\vert
0_{A},0_{a},\uparrow ,\downarrow \right\rangle -\left\vert
0_{A},0_{a},\downarrow ,\uparrow \right\rangle \right) ,  \notag \\
\left\vert T_{0}\right\rangle &=&\frac{1}{\sqrt{2}}\left( \left\vert
0_{A},0_{a},\uparrow ,\downarrow \right\rangle +\left\vert
0_{A},0_{a},\downarrow ,\uparrow \right\rangle \right) , \\
\left\vert T_{-1}^{A}\right\rangle &=&\left\vert 1_{A},0_{a},\downarrow
,\downarrow \right\rangle ,\left\vert T_{-1}^{a}\right\rangle =\left\vert
0_{A},1_{a},\downarrow ,\downarrow \right\rangle .  \notag
\end{eqnarray}%
The eigenstates of the Hamiltonian in this subspace are%
\begin{eqnarray}
\left\vert \Phi _{A}^{+}\right\rangle &=&\sin \frac{\theta _{A}}{2}%
\left\vert T_{-1}^{A}\right\rangle +\cos \frac{\theta _{A}}{2}\left\vert
T_{0}\right\rangle , \\
\left\vert \Phi _{A}^{-}\right\rangle &=&\cos \frac{\theta _{A}}{2}%
\left\vert T_{-1}^{A}\right\rangle -\sin \frac{\theta _{A}}{2}\left\vert
T_{0}\right\rangle , \\
\left\vert \Phi _{a}^{+}\right\rangle &=&\sin \frac{\theta _{a}}{2}%
\left\vert T_{-1}^{a}\right\rangle +\cos \frac{\theta _{a}}{2}\left\vert
S\right\rangle , \\
\left\vert \Phi _{a}^{-}\right\rangle &=&\cos \frac{\theta _{a}}{2}%
\left\vert T_{-1}^{a}\right\rangle -\sin \frac{\theta _{a}}{2}\left\vert
S\right\rangle
\end{eqnarray}%
and the corresponding eigenvalues are:%
\begin{eqnarray}
E_{A}^{\pm } &=&-\frac{\Delta _{A}}{2}\pm \sqrt{\left( \frac{\Delta _{A}}{2}%
\right) ^{2}+2g_{A}^{2}},  \label{ExactEnergy1} \\
E_{a}^{\pm } &=&-\frac{\Delta _{a}}{2}\pm \sqrt{\left( \frac{\Delta _{a}}{2}%
\right) ^{2}+2g_{a}^{2}}.  \label{ExactEnergy}
\end{eqnarray}%
Here, the zero point energies of the four boson modes are omitted. $\theta
_{A,a}$ is defined as
\begin{equation}
\tan \theta _{A,a}=\frac{2\sqrt{2}g_{A,a}}{\Delta _{A,a}}
\end{equation}%
where $\theta _{A,a}\in \left[ 0,\pi \right] $, and $\Delta _{A,a}=\omega
_{z}-\omega _{A,a}$.

With the help of these exact eigenenergies and the eigenstates obtained
above, we can calculate the FIR absorption spectra analytically. To this
end, we consider the time-dependent Hamiltonian of the system in an optical
field is%
\begin{equation}
\tilde{H}(t)=H+H^{\prime }e^{-i\omega t}
\end{equation}%
where $H^{\prime }$ is the time dependent term induced by the classical
optical field \cite{Chakraborty2005}%
\begin{equation}
H^{\prime }\propto -\sum_{i=1,2}\left[ \frac{ea}{m}\boldsymbol{\epsilon }%
\cdot \left( \mathbf{p}_{i}+e\mathbf{A}_{i}\right) +\frac{\alpha ea}{\hbar }%
\left( \mathbf{\sigma }_{i}\times \boldsymbol{\epsilon }\right) _{z}\right]
\end{equation}%
where $a$ is the radiation field amplitude, and $\boldsymbol{\epsilon }%
=\left( \hat{x}-i\hat{y}\right) /\sqrt{2}$ is the polarization vectors for
circular polarized light.

The absorption coefficient is calculated according to the Fermi golden rule%
\cite{Spectrum}\cite{SpectrumHuhui}:%
\begin{equation}
\alpha \left( \omega \right) \propto \omega \sum\limits_{f}\left\vert
\left\langle f|H^{\prime }|i\right\rangle \right\vert ^{2}\delta \left(
\omega -\omega _{fi}\right) ,
\end{equation}%
Here, $\hbar \omega _{fi}=E_{f}-E_{i}$, $\left\vert i\right\rangle
=\left\vert GS\right\rangle $ is the ground state, and $\left\vert
f\right\rangle $ stands for the excited states. We will focus on the four
lowest excited states $\left\vert \Phi _{A}^{\pm }\right\rangle $ and $%
\left\vert \Phi _{a}^{\pm }\right\rangle $. According to Eqs. (\ref%
{bosonOperators}) and (\ref{bosonOperators2}), the perturbation $H^{\prime }$
is obtained as%
\begin{equation}
H^{\prime }=\hbar \Omega \sqrt{\frac{\omega _{0}}{\Omega }}\left( 1-\frac{%
\omega _{c}}{\Omega }\right) A^{+}+\alpha \sqrt{\frac{2m_{0}\omega _{0}}{%
\hbar }}\left( \sigma _{1+}+\sigma _{2+}\right)  \label{OpticallyHami}
\end{equation}

Note that we have omitted the terms related to the $B$ modes, which is not
involved in the initial and final states we are considering and thus does
not contribute to the absorption coefficient. Our analytical results give
the FIR spectra of obvious physical meanings. From Eq.(\ref{OpticallyHami})
above, we find that the first term is spin independent, and it provides an
CM angular excitation. This CM\ angular excitation, which is the consequence
of Kohn theorem\cite{KohnTheorem}, exists even in the absence of SOI. The
second term, which is spin dependent, is due to the presence of SOI. This
term contributes an excitation of the two spins \textit{symmetrically}, i.e.
$\left\vert \downarrow \downarrow \right\rangle \longmapsto \left\vert
\uparrow \downarrow \right\rangle +\left\vert \downarrow \uparrow
\right\rangle $. Thus, $H^{\prime }$ only couples the ground state $%
\left\vert GS\right\rangle $ with the CM\ excitation states $\left\vert \Phi
_{A}^{\pm }\right\rangle $, and $\left\vert \Phi _{a}^{\pm }\right\rangle $
are inactive in this case. The matrix elements of $\left\langle \Phi
_{A}^{\pm }|H^{\prime }|GS\right\rangle $ are calculated, and the FIR
spectra are shown in Fig.\ref{EnergySpctra}(b).

\bigskip

\section{Quantum entanglement in large detuning limit}

Due to the linearly increasing of the dimension of the invariance subspace $%
V^{\left( N\right) }$ with respect to $N$, it is difficult to obtain a
compact solution of the eigenvalue problem for $N\geq 1$. Instead of the
exact solution, the approximate solution by perturbation theory will be
given in this section. In the perturbation available regime, we derive an
effective transverse spin-spin interaction Hamiltonian. This Hamiltonian can
induce a two-qubit logic gate, and can be used to produce a controllable
quantum entanglement.

We summarize the Hamiltonian obtained%
\begin{eqnarray}
H_{0} &=&H_{\text{orbit}}+\frac{1}{2}\hbar \omega _{z}\left( \sigma
_{1z}+\sigma _{2z}\right) , \\
H_{1} &=&g_{A}\cdot A\left( \sigma _{1+}+\sigma _{2+}\right)  \notag \\
&+&g_{a}\cdot a\left( \sigma _{1+}-\sigma _{2+}\right) +h.c.
\label{FullHami}
\end{eqnarray}%
and then consider its reduction in the large detuning limit, i.e. $\Delta
_{A,a}>>g_{A,a}$. In this limit, we perform the Fr{\"{o}h}lich transform
with the operator
\begin{equation}
S=\left( \frac{g_{A}}{\Delta _{A}}A^{+}\left( \sigma _{1-}+\sigma
_{2-}\right) +\frac{g_{a}}{\Delta _{a}}a^{+}\left( \sigma _{1-}-\sigma
_{2-}\right) \right) -h.c.
\end{equation}%
and the effective Hamiltonian $\exp (-S)H\exp (S)$ is calculated up to the
second order as:%
\begin{equation}
H_{S}\simeq H_{0}+\frac{1}{2}\left[ H_{1},S\right] .  \label{Heffshort}
\end{equation}%
Here, the second term in the r.h.s can be written explicitly
\begin{widetext}
\begin{align}
& \frac{1}{2}\left[ H_{1},S\right] =\hbar \xi \left( \sigma _{1+}\sigma
_{2-}+\sigma _{1-}\sigma _{2+}\right) +\frac{g_{A}g_{a}}{2}\left( \frac{1}{%
\Delta _{A}}+\frac{1}{\Delta _{a}}\right) \left( A^{+}a+a^{+}A\right) \left(
\sigma _{1z}-\sigma _{2z}\right)   \notag \\
& +\left[ \frac{g_{A}^{2}}{\Delta _{A}}\left( A^{+}A+1/2\right) +\frac{%
g_{a}^{2}}{\Delta _{a}}\left( a^{+}a+1/2\right) \right] \left( \sigma
_{1z}+\sigma _{2z}\right) +\left( \frac{g_{A}^{2}}{\Delta _{A}}+\frac{%
g_{a}^{2}}{\Delta _{a}}\right)   \label{EffectiveHami1}
\end{align}%
\end{widetext}

Note that the first term of Eq.(\ref{EffectiveHami1}) is the effective
transverse spin-spin coupling induced by SOI. The effective coupling strength%
\begin{equation}
\hbar \xi =g_{A}^{2}/\Delta _{A}-g_{a}^{2}/\Delta _{a}
\end{equation}%
depends on (i) the SOI strength $\alpha $ (see Eqs.(\ref{CouplingStrengtha})
and (\ref{CouplingStrengthA})), and (ii) the frequency difference between $%
\omega _{A}$ and $\omega _{a}$, which is the consequence of the Coulomb
interaction. We note that, in the subspace $V^{\left( 0\right) }$, the
second term of Eq.(\ref{EffectiveHami1}) vanishes, and the remaining terms
commute with the total spin $\boldsymbol{\sigma }^{2}=\left( \boldsymbol{%
\sigma }_{1}+\boldsymbol{\sigma }_{2}\right) ^{2}$. Thus we can denote the
eigenstates by spin singlet $\left\vert S\right\rangle =\left( \left\vert
\uparrow \downarrow \right\rangle -\left\vert \downarrow \uparrow
\right\rangle \right) /\sqrt{2}$ and triplet $\left\vert T_{0}\right\rangle
=\left( \left\vert \uparrow \downarrow \right\rangle +\left\vert \downarrow
\uparrow \right\rangle \right) /\sqrt{2}$, $\left\vert T_{1}\right\rangle
=\left\vert \uparrow \uparrow \right\rangle $, and $\left\vert
T_{-1}\right\rangle =\left\vert \downarrow \downarrow \right\rangle $ in the
large detuning limit in subspace $V^{\left( 0\right) }$. Diagonalizing the
Hamiltonian (\ref{EffectiveHami1}), we obtain the eigenenergies
\begin{eqnarray}
E_{\left\vert T_{-1}^{A}\right\rangle } &=&-\Delta _{A}-\frac{2g_{A}^{2}}{%
\Delta _{A}},  \label{perturbEnergy1} \\
E_{\left\vert T_{-1}^{a}\right\rangle } &=&-\Delta _{a}-\frac{2g_{a}^{2}}{%
\Delta _{a}}, \\
E_{\left\vert T_{0}\right\rangle } &=&\frac{2g_{A}^{2}}{\Delta _{A}}, \\
E_{\left\vert S\right\rangle } &=&\frac{2g_{a}^{2}}{\Delta _{a}}
\label{perturbEnergy}
\end{eqnarray}%
for the different spin states corresponding to the exact solutions in Eq.(%
\ref{ExactEnergy}). Here, we also omit the zero point energies.

The above energies are drawn as the function of magnetic field in Fig.\ref%
{perturb}(c) in comparison with the exact solution. The exact consistency of
the two solutions in the large detuning regime confirms the validity of our
perturbation treatment.
\begin{figure}[tbp]
\includegraphics[bb=15 51 550 775, width=7 cm, clip]{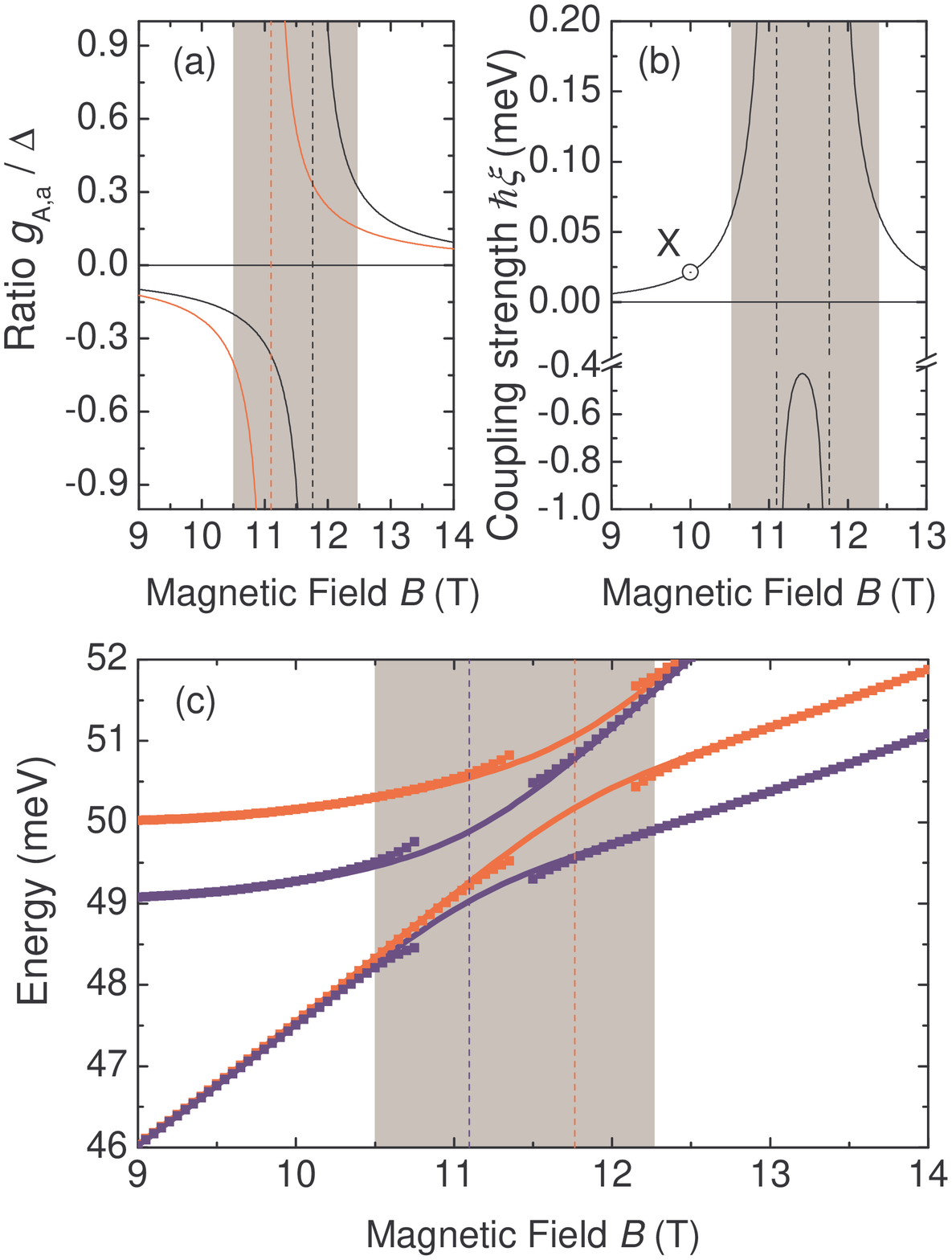}
\caption{{}(a) The ratio $g_{A}/\Delta _{A}$ (black line) and, $g_{a}/\Delta
_{a}$ (red line). The perturbation treatment is valid when both $%
g_{A}/\Delta _{A}$ and $g_{a}/\Delta _{a}$ less than unity. (b) The
effective transverse spin coupling strength induced by SOC. (c) A comparison
of eigenenergies between exact (Eqs.(\protect\ref{ExactEnergy1}) and (%
\protect\ref{ExactEnergy}), lines) and perturbative(Eqs.(\protect\ref%
{perturbEnergy1}) \symbol{126}(\protect\ref{perturbEnergy}), scattered dots)
solutions in the large detuning regime. Our perturbation treatment is valid
in the unshaded region.}
\label{perturb}
\end{figure}
Fig.\ref{perturb}(a) shows effective spin coupling strength induced by SOI
in our model. In the perturbation valid regime, for example at $B=10T$ (the
point denoted by X), we have the spin coupling strength $\hbar \xi \simeq
20\mu eV$. This value is comparable with that in the proposal of the
spin-spin coupling induced by the by electromagnetic field in cavity\cite%
{Imamoglu}. Thus, it is feasible to realize the two-qubit gate operation
during the long coherence time of conduction band electrons.

The spin transverse coupling described by the first term in the r.h.s. of
Eq.(\ref{EffectiveHami1}) generates an ideal $\sqrt{i\text{SWAP}}$ gate\cite%
{iSWAPgate}\cite{DLoss} at a specific time $t_{0}=\pi /4\xi $. On the other
hand, it is worthy to notice that the unitary transformation $\exp (-S)H\exp
(S)$ may induce an excitation of the boson modes, and then causes gate error
during the time evolution of the spin states. 
\begin{figure}[tbp]
\includegraphics[bb=15 60 550 775, width=7 cm, clip]{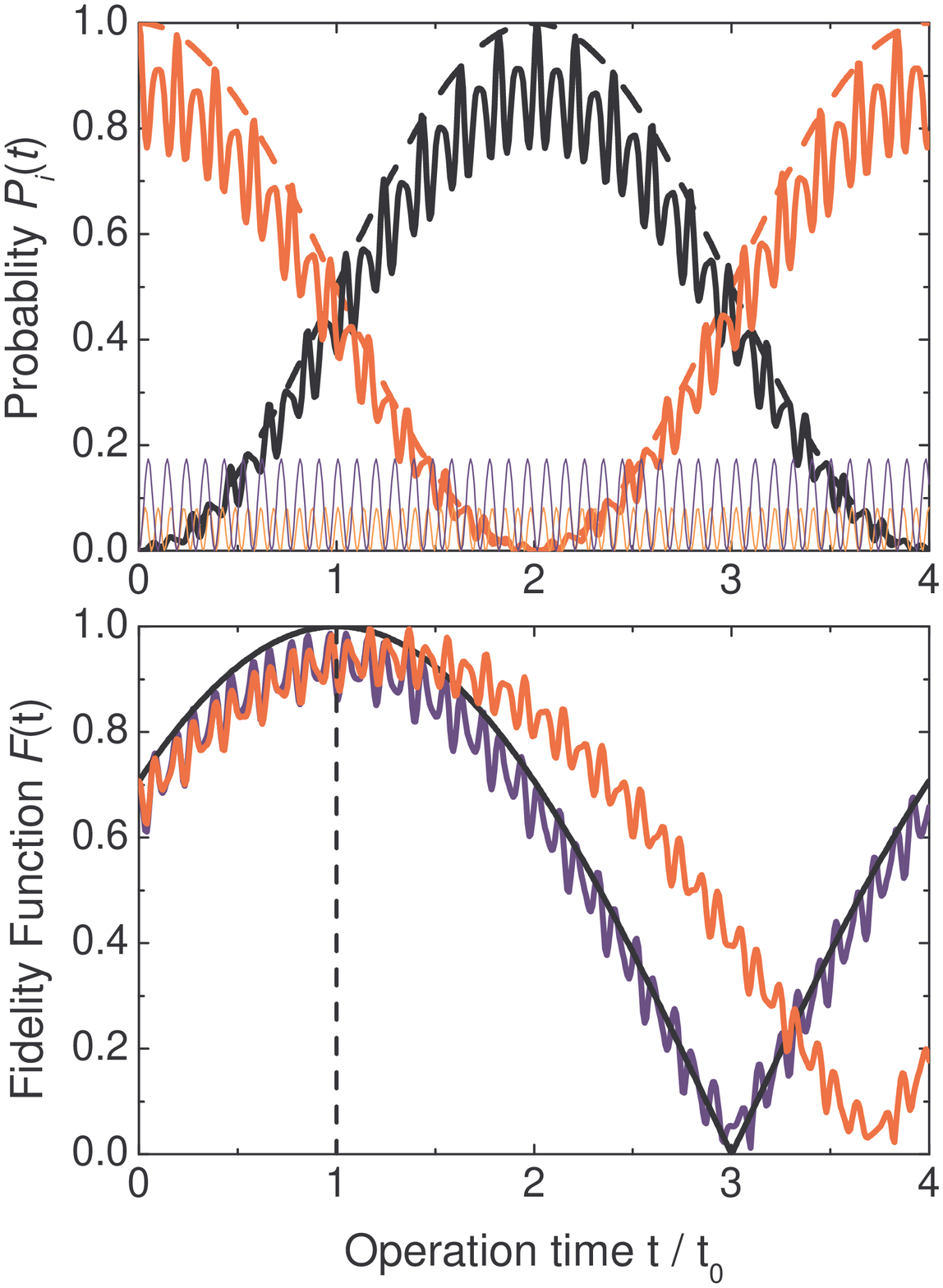}
\caption{(a)Probability $P_{i}\left( t\right) $ as a function of $t$. Black,
red, blue, and orange solid lines represent the probabilities of $\left\vert
\uparrow \downarrow ,0_{A},0_{a}\right\rangle $, $\left\vert \downarrow
\uparrow ,0_{A},0_{a}\right\rangle $, $\left\vert \downarrow \downarrow
,1_{A},0_{a}\right\rangle $, and $\left\vert \downarrow \downarrow
,0_{A},1_{a}\right\rangle $ respectively. The dashed lines are the
probabilities generated by the standard $\protect\sqrt{i\text{SWAP}}$ gate.
(b)Gate fidelity as a function of $t$. The black line indicates the ideal $%
\protect\sqrt{i\text{SWAP}}$ operation. The blue and red lines represent the
fidelity function generated by the second order approximation Hamiltonian
and the original Hamiltonian (see text) respectively.}
\label{fidelityplot}
\end{figure}
To explicitly show the unwanted boson modes excitation and to examine the
reliability of our model, we consider the time evolution of the initial
state $\left\vert \psi \left( 0\right) \right\rangle =\left\vert \downarrow
\uparrow ,0_{A},0_{a}\right\rangle $:%
\begin{equation*}
\left\vert \psi \left( t\right) \right\rangle =U\left( t\right) \left\vert
\psi \left( 0\right) \right\rangle ,
\end{equation*}%
where%
\begin{equation*}
U\left( t\right) =\exp \left[ it\left( e^{S}H_{S}e^{-S}\right) /\hbar \right]
=e^{S}\exp \left( itH_{S}/\hbar \right) e^{-S}.
\end{equation*}%
Notice that $\left\vert \psi \left( 0\right) \right\rangle $ belongs to the
subspace $V^{\left( 0\right) }$, and the time evolution will be restricted
in this subspace. Fig\ref{fidelityplot}(a) gives the probabilities $%
P_{i}\left( t\right) =\left\vert \left\langle i|\psi \left( t\right)
\right\rangle \right\vert ^{2}$ of finding state $\left\vert i\right\rangle $
at time $t$, where $\left\vert i\right\rangle $ stands for the four bases of
the subspace $V^{\left( 0\right) }$: $\left\vert \uparrow \downarrow
,0_{A},0_{a}\right\rangle $, $\left\vert \downarrow \uparrow
,0_{A},0_{a}\right\rangle $, $\left\vert \downarrow \downarrow
,1_{A},0_{a}\right\rangle $, and $\left\vert \downarrow \downarrow
,0_{A},1_{a}\right\rangle $ respectively. Besides the states $\left\vert
\uparrow \downarrow ,0_{A},0_{a}\right\rangle $ and $\left\vert \downarrow
\uparrow ,0_{A},0_{a}\right\rangle $, the unwanted boson modes excitation
states $\left\vert \downarrow \downarrow ,1_{A},0_{a}\right\rangle $, and $%
\left\vert \downarrow \downarrow ,0_{A},1_{a}\right\rangle $ also have
nonzero populations. This populations will induce the gate error.
Furthermore, we calculate the fidelity function defined by%
\begin{equation}
F\left( t\right) =\left\vert \left\langle \psi \left( 0\right) \left\vert U_{%
\sqrt{i\text{SWAP}}}^{+}U\left( t\right) \right\vert \psi \left( 0\right)
\right\rangle \right\vert ,
\end{equation}%
where $U_{\sqrt{i\text{SWAP}}}$ is the ideal $\sqrt{i\text{SWAP}}$ gate
operator. We notice that the fidelity function $F\left( t\right) $ reaches
its maximum slightly less than unity at $t=t_{0}$, and the high-frequency
oscillations appear due to the boson modes excitation mentioned above. We
also examine the fidelity function generated by the original Hamiltonian (%
\ref{FullHami}), i.e. $U\left( t\right) =\exp \left[ -it\left(
H_{0}+H_{1}\right) /\hbar \right] $. In this case the original Hamiltonian (%
\ref{FullHami}) gives a high fidelity $F\left( t^{\prime }\right) \simeq 1$
at a different time $t^{\prime }\simeq 1.2t_{0}$, which can be regarded as
the higher order correction in comparison to the approximate Hamiltonian (%
\ref{Heffshort}). Finally, Fig.\ref{fidelityplot} shows that spin
entanglement can be created in this system by adjusting the operation time.


\section{Conclusion}

In this paper, we have considered a system of two vertically CQDs each
containing an electron in the presence of Rashba type SOI. We theoretically
demonstrate that it is possible to create spin entanglement in this kind of
system by using the SO coupling. With the large interdot separation case,
the Coulomb interaction between the two electrons is approximately expressed
in a quadratic form. And then two-boson-two-spin interacting model is
derived in the RWA from the original Hamiltonian. We give the exact solution
of the low excitation states analytically. This solution helps us reveal the
physics under the FIR spectra near the resonant point. Perturbation
treatment in the large detuning case shows that, similar to the quantum dot
embedded in an optical cavity, the orbital freedoms play a role of quantized
data bus via the Coulomb interaction and SOI in this system. An effective
Hamiltonian of spin-spin interaction is obtained in the perturbation regime
by eliminating the orbital freedoms. This Hamiltonian provides a tow-qubit
operation, which is essential in quantum information processing.

Finally, we would like to point out that using the effective inter-spin
coupling to create spin entanglement is feasible to be controlled and
measured. From the discussion above, we know that the effective spin
coupling strength can be controlled by external magnetic field. On the other
hand, the tuneable strength of SOI $\alpha $ \cite{Debald2005}, in
principle, also enable\ us to switch on and off the effective inter-spin
coupling by external gates conveniently. To probe the quantum entanglement
of spin system, similar method in the protocol proposed in ref. \cite%
{Measure-Loss} can be used, where the information stored in the spin degrees
of freedom is converted to the charge states, and then the charge states can
be detected.

This work is funded by NSFC with grant Nos. 90203018, 10474104, 60433050,
10374057, 10574077 and NFRPC with Nos. 2001CB309310, 2005CB724508,
2005CB623606.

\end{document}